\documentclass[12pt,a4paper]{article}
\usepackage[intlimits,centertags]{amsmath}
\usepackage{amssymb,amsfonts}
\usepackage{bbm}
\usepackage{mathrsfs}
\usepackage[vcentermath]{youngtab}
\usepackage{cite}
\usepackage[scale={0.75,0.81}]{geometry}

\newlength{\textlength}
\newlength{\overlinelength}
\newcommand{\ol}[2][.625]{%
   \settowidth{\textlength}{\ensuremath{#2}}%
   \setlength{\overlinelength}{3pt}%
   \addtolength{\overlinelength}{0.4\textlength}%
   \makebox[\textlength][s]{\ensuremath{#2}}%
   \hspace{-.5\textlength}\hspace{-\overlinelength}\hspace{#1\overlinelength}
   \overline{%
      \makebox[\overlinelength][c]{%
         \vphantom{\ensuremath{#2}}
      }
   }
   \hspace{-#1\overlinelength}\hspace{.5\textlength}
}

\newcommand{\I}{\ensuremath{\text{i}}}
\newcommand{\D}{\ensuremath{\text{d}}}
\newcommand{\Lambdabs}{\ensuremath{\bar{\scriptstyle \Lambda}}}
\newcommand{\Lambdab}{\bar{\Lambda}}
\newcommand{\lambdab}{\bar{\lambda}}
\newcommand{\thetab}{\ensuremath{\bar{\theta}}}
\newcommand{\varepsilonb}{\ensuremath{\bar{\varepsilon}}}
\newcommand{\Phib}{\ensuremath{\ol[.49]{\Phi}}}
\newcommand{\Psib}{\ensuremath{\ol[.49]{\Psi}}}
\newcommand{\psib}{\ensuremath{\ol[.49]{\psi}}}
\newcommand{\xib}{\ensuremath{\ol[.49]{\xi}}}
\newcommand{\epsilonb}{\ensuremath{\bar{\epsilon}}}
\newcommand{\phib}{\ensuremath{\ol[.49]{\phi}}}
\newcommand{\chib}{\ensuremath{\ol[.49]{\chi}}}
\newcommand{\etab}{\ensuremath{\ol[.49]{\eta}}}
\newcommand{\Mh}{{\widehat{M}}}
\newcommand{\Nh}{{\widehat{N}}}
\newcommand{\Ch}{{\widehat{C}}}
\newcommand{\Fb}{\ensuremath{\ol{F}}}
\newcommand{\Db}{\ensuremath{\ol{D}}}
\newcommand{\sigmat}{\ensuremath{\widetilde{\sigma}}}
\newcommand{\Gammah}{\ensuremath{\widehat{\Gamma}}}
\newcommand{\loongrightarrow}{\ensuremath{\xrightarrow{\mspace{68mu}}}}
\newcommand{\alphad}{{\ensuremath{\dot{\alpha}}}}
\newcommand{\betad}{{\ensuremath{\dot{\beta}}}}

\DeclareMathOperator{\cons}{const}
\DeclareMathOperator{\tr}{tr}
\DeclareMathOperator{\diag}{diag}

\allowdisplaybreaks[2]
\numberwithin{equation}{section}

\begin{document}
  \title{Seven-Dimensional Super-Yang--Mills Theory in $\mathcal{N}=1$ Superfields}
  \author{Christoph L\"{u}deling\\[2mm]
    {\normalsize\itshape Bethe Center for Theoretical Physics and Physikalisches Institut der
      Universit\"at Bonn,}\\{\normalsize\itshape Bonn, Germany}\\ 
    {\small\ttfamily luedeling@th.physik.uni-bonn.de}}

  \maketitle
    \begin{abstract}
      We give a gauge-covariant formulation of seven-dimensional super-Yang--Mills theory in terms of
      $\mathcal{N}=1$ superfields. Furthermore, we show that five and seven dimensions are the only cases where such a
      formulation exists by analysing the interplay of Lorentz and $R$ symmetries. The action is expressed in terms of
      field strengths and a Chern--Simons-like superpotential. Each term is manifestly $\mathcal{N}=1$ supersymmetric,
      Lorentz invariant in four dimensions and gauge invariant under superfield gauge transformations, including 
      those that do not preserve Wess--Zumino gauge.   
    \end{abstract}

  \section{Introduction}
    Superfields \cite{Salam:1974yz} are a very convenient tool for model building in $\mathcal{N}=1$
    supersymmetric\footnote{Throughout this 
    paper, we use the four-dimensional $\mathcal{N}$, i.e.\ $\mathcal{N}=1$ corresponds to four supercharges.}
    theories. For higher supersymmetries, however, superspace formulations
    \cite{Lindstrom:1989ne} are rather less convenient and have not been used much. This also
    applies to higher-dimensional  theories, which from the four-dimensional perspective correspond to  $\mathcal{N}=2$ or
    $\mathcal{N}=4$ supersymmetry.  There have been various reformulations of higher-dimensional
    supersymmetric theories in terms of $\mathcal{N}=1$ superfields, starting
    with~\cite{Marcus:1983wb} for the ten-dimensional theory. There was renewed interest in the
    subject with the advent of higher-dimensional (orbifold) field theory models since around 2000
    \cite{ArkaniHamed:2001tb,Marti:2001iw,Belyaev:2006jg}. The general idea is that the fields of any theory with
    higher supersymmetry still form multiplets under an $\mathcal{N}=1$ subset of the symmetry, and
    hence fit into $\mathcal{N}=1$ superfields in terms of which one can write the action. For
    example, five- or six-dimensional models 
    correspond to $\mathcal{N}=2$ supersymmetry, where the vector multiplet gives rise to a vector
    and a chiral superfield, while the hypermultiplet gives two chiral superfields. In seven to ten
    dimensions, there is only the vector multiplet (in rigid supersymmetry, i.e.\ excluding
    supergravity), which leads to one vector and three chiral superfields. This idea has also been
    extended to five-dimensional supergravity~\cite{Paccetti:2004ri} (see also \cite{Linch:2002wg} for the linearised
    case). 

    It might seem that in this approach one looses a lot of manifest symmetry.
    However, 
    usually one is interested in $\mathcal{N}=1$ models in four dimensions anyway, and the remaining supersymmetry is
    broken by the process of compactification, such as orbifold twists, nontrivial holonomy or intersecting branes.
    The $\mathcal{N}=1$ superfield approach then suitable for dealing with parts of the model that do not respect the
    full supersymmetry, but only a subset, such as localised matter.

    For gauge theories, however, there is a common problem in these approaches: The action is in
    general not formulated in terms of covariant derivatives and fields strengths only, but rather
    contains explicit factors of the vector superfield $V$ and/or partial derivatives in the extra
    dimensions. This means that the invariance of the 
    action under gauge transformations that do not respect Wess--Zumino (WZ) gauge is not guaranteed and
    has to be enforced by a Wess--Zumino--Witten (WZW) like term \cite{Marcus:1983wb} of the form
    \begin{align}
      \left(\partial V\right) \frac{\sinh L_V -L_V}{L_V^2} \left(\partial V\right)&=\left(\partial
        V\right) \left(\frac{L_V}{3!} + \frac{L_V^3}{5!}+\dotsm\right) \left(\partial V\right)
    \end{align}
    with the Lie bracket $L_V X=\left[V,X\right]$. Here $\partial V$ denotes some partial derivatives of $V$ in the
    internal dimensions. This term vanishes in WZ gauge since already the
    first term in the series is $\mathcal{O}\!\left(V^3\right)$, but for a general form of $V$,
    including a $\theta=\thetab=0$ component, the series is indeed infinite. 

    The situation has been significantly simplified in five dimensions by Hebecker
    \cite{Hebecker:2001ke} by giving a fully gauge covariant description. A key ingredient was the
    introduction of a covariant derivative in the extra dimension, which allowed to define a field
    strength analogous to the standard $W_\alpha$, in terms of which the action can be easily
    formulated. The aim of this paper is to extend this formulation to other dimensions. This will,
    however, turn out to be possible only for the seven-dimensional case, and possibly for
    six-dimensional $\mathcal{N}=4$ super-Yang--Mills theory, as can be seen by
    considering the respective $R$~symmetries.  

    The paper is organised as follows: In Section~\ref{sec:two}, we will review the covariant
    description of~\cite{Hebecker:2001ke} and discuss the possibilities of extending the method to
    other dimensions. In Section~\ref{sec:comp}, we will then discuss the component form of the
    seven-dimensional theory, and the decomposition in terms of four-dimensional degrees of freedom.
    The superfield embedding and action are given in Section~\ref{sec:sf}. We will finally conclude in
    Section~\ref{sec:conclusions} and mention some potential applications. Finally, in the appendices we present some
    details on the reduction of the spinors and the supersymmetry transformations.

  \section{Covariant Formulations\label{sec:two}}
    \subsection{Five Dimensions}
      A covariant formulation of five-dimensional super-Yang--Mills theory was given by He\-becker \cite{Hebecker:2001ke},
      and we will briefly review the central point. In five dimensions, the off-shell theory is known
      \cite{Mirabelli:1997aj}. It involves the gauge vector $A_M$, 
      a scalar $B$ and a symplectic Majorana spinor $\lambda_I$ as dynamical fields, as well as
      an $SU(2)_R$ triplet $X^i$ of auxiliary fields\footnote{For consistency with later
        sections, our notation differs from \cite{Hebecker:2001ke}.}.  
      They form an off-shell representation of the SUSY algebra under the transformations
      \begin{subequations}
        \begin{align}
          \delta A_M&=\I \varepsilonb^I \gamma_M\lambda_I\,,\\
          \delta B&= \I \varepsilonb^I\lambda_I\,,\\
          \delta \lambda_I&=\gamma^{MN}F_{MN} \,\varepsilon_I +\gamma^M D_M B \,\varepsilon_I +\I
          X_i\left(\sigma^i\right)^J_I \varepsilon_J\,,\\
          \delta X^i&=\varepsilonb^I\left(\sigma^i\right)^J_I\gamma^M D_M\lambda_J
          +\I\left[B,\lambda_J\right] \left(\sigma^i\right)^J_I \varepsilonb^I\,.
        \end{align}
      \end{subequations}
      Here also the variation parameter $\varepsilon_I$ is a symplectic Majorana spinor, which
      corresponds to two 4D Weyl spinors. The key observation is now that under a SUSY
      transformation generated by one of these Weyl spinors, one can identify a vector and a chiral
      superfield,
      \begin{align}
        V&=-\theta \sigma^\mu \thetab A_\mu +\I\theta^2\thetab\lambdab -\I \thetab^2 \theta\lambda
        +\frac{1}{2} \theta^4 \left(X^3-D_5 B\right)\,,\\
        \Phi&= A_5 +\I B +2\theta\psi +\theta^2 \left(X^1+\I X^2\right)\,.
      \end{align}
      Here $\lambda$ and $\psi$ arise from a suitable decomposition of the 5D spinor $\lambda_I$
      into Weyl spinors. Under gauge transformations with superfield parameter $\Lambda$, these
      superfields transform as 
      \begin{align}
        e^{2V}&\loongrightarrow e^{-\I\Lambdabs} e^{2V} e^{\I\Lambda}\,, & \Phi&\loongrightarrow
        e^{-\I\Lambda} \left(\Phi-\I\partial_5 \right) e^{\I\Lambda}\,.
      \end{align}
      $V$ is given in Wess--Zumino gauge, and consequently, gauge and supersymmetry transformations
      mix (i.e.\ a SUSY transformation requires a compensating gauge transformation to return to WZ
      gauge). The same now applies to the chiral field $\Phi$.     

      The particular form of $\Phi$ allows to define a derivative that is covariant with respect to
      supersymmetry and gauge symmetry,
      \begin{align}
        \nabla=\partial_5+\I\Phi\,.
      \end{align}      
      With this derivative, one can define a covariantly transforming ``extra-dimensional field
      strength'' 
      \begin{align}
        Z&=e^{-2V} \nabla e^{2V}\,, &Z&\loongrightarrow e^{-\I\Lambda} Ze^{\I\Lambda}\,. 
      \end{align}
      $Z$, together with the usual field strength $W_\alpha$, elegantly reproduces the 5D component Lagrangean:
      \begin{align}
        \begin{split}
          \mathscr{L}_5&=\frac{1}{4}\int \D^4 \theta \tr Z^2 + \frac{1}{4}\left(\int \D^2\theta \tr W^\alpha
            W_\alpha + \text{H.c.}\right) \\
          &=-\frac{1}{4} F_{MN}F^{MN} -\frac{1}{2} D_M\, B D^M B -\frac{\I}{2}\lambdab^I
          \Gamma^M D_M \lambda_I +\frac{1}{2} X^i X_i   +\frac{1}{2}
          \lambdab^I\left[B,\lambda_I\right] \,.
        \end{split}
      \end{align}
      The superfield action is constructed from covariant quantities only and does not contain
      explicit factors of $V$, so it is manifestly gauge invariant under arbitrary gauge
      transformations, in particular under those which do not maintain Wess--Zumino gauge.

    \subsection{\boldmath$R$ Symmetries and General Dimensions}
      The approach outlined in the previous Section cannot be generalised to arbitrary dimensions.
      This can be seem from a symmetry argument: When expressed in four-dimensional degrees of
      freedom, $4+d$-dimensional super-Yang--Mills theory corresponds to a $\mathcal{N}=2$ or
      $\mathcal{N}=4$ theory, which would have an $R$ symmetry $SU(2)$ or $SU(4)$, respectively.
      However, there is extra structure which reduces the $R$ symmetry: Besides the vector field, the theory will
      contain two or six 
      scalar fields, which separate into $d$ extra-dimensional vector components, which transform under $SO(d)$ and a
      number of $\left(4+d\right)$-dimensional scalars which transform under $SO(2-d)$ or $SO(6-d)$, respectively.
      The fermionic sector will likewise contain two or four Weyl
      fermions obtained from the decomposition of the gaugino, which form a fundamental
      representation of $SU(2)$ or $SU(4)$. A superfield approach such as above singles out one of
      the Weyl fermions to be the 4D gaugino, hence it breaks the manifest $R$~symmetry from $SU(2)$
      to nothing or $SU(4)\to SU(3)$, respectively.

      Hence, we have different symmetries in the scalar and the fermionic
      sector, unless $d=1$, where there is no such symmetry, or $d=3$, where we can embed the
      scalars' diagonal $SO(3)$ in the fermions' $SU(3)$. In a more pedestrian view, we want to form chiral
      multiplets whose scalar components are of the form $A+\I B$, where $A$ is an extra-dimensional vector component
      and $B$ is a true scalar. The number of $A$'s and $B$'s coincides only in five and seven dimensions. 
      
      This argument does furthermore suggest that a similar description is possible for
      six-dimensional $\mathcal{N}=2$ super-Yang--Mills theory, where the fermions again have an
      $SU(3)$ symmetry into which the geometric $SO(2)$ can be embedded. Then one of the three chiral multiplets will be
      a pure adjoint matter multiplet.

  \section{Component Lagrangean\label{sec:comp}}
    Minimal supersymmetry in seven to ten dimensions has 16 supercharges, i.e.\ it is $\mathcal{N}=4$
    from the four-dimensional point of view. Hence the only multiplet with spins not greater than
    one is the Yang-Mills multiplet, but there are no matter multiplets. The precise field content and action
    in seven dimensions can be 
    derived from the super-Yang--Mills theory in ten dimensions, where the theory has the ``minimal''
    field content, i.e.\ just the gauge field $A_\Mh$ and a Majorana--Weyl spinor $\Xi$ in the
    adjoint representation of the gauge group $G$. The Lagrangean in ten dimensions is
    \begin{align}\label{eq:L10}
      \mathscr{L}_{10}&= -\frac{1}{4} F_{\Mh\Nh} F^{\Mh\Nh}- \frac{\I}{2} \overline{\Xi}\Gamma^\Mh D_\Mh
      \Xi \,,
    \end{align}
    the SUSY transformations are
    \begin{align}
      \delta A_\Mh &= \frac{\I}{2}\epsilonb\Gamma_\Mh \Xi\,, & \delta \Xi &=
      -\frac{1}{4}F_{\Mh\Nh}\Gamma^{\Mh\Nh}\epsilon\,.
    \end{align}
    Here and in the following, we set the coupling constant to $g=1$. It can always be restored by
    dimensional arguments.
    
    The reduction to seven dimensions is basically straightforward. The only subtlety lies in the
    different types of spinors in ten and seven dimensions \cite{Strathdee:1986jr}: The
    seven-dimensional superalgebra has an $SU(2)_R$ symmetry, even for minimal supersymmetry. Some
    details about the reduction are given in Appendix~\ref{sec:10to7}.
    The fields of the seven-dimensional theory are thus a vector $A_M$ which is a singlet
    under the $R$~symmetry, a triplet of scalars $B_i$ and doublet of spinors $\Psi_I$ which
    satisfy a symplectic Majorana condition,
    \begin{align}
      \Psi_I&=\varepsilon_{IJ} C \left(\Psib^J\right)^T\,.
    \end{align}
    
    The Lagrangean is 
    \begin{align}
      \begin{split}\label{eq:Lsym7d}
        \mathscr{L}_7&= -\frac{1}{4}\tr F_{MN} F^{MN} -\frac{1}{2} \tr D_M B_i D^M B^i +\frac{1}{4}
        \tr \left[B_i,B_j\right]  \left[B^i,B^j\right] \\
        &\quad - \frac{\I}{2}\tr \Psib^I \Gamma^M D_M \Psi_I -\frac{\I}{2} \tr
        \Psib^I \left[B_i \left(\sigma^i\right)_I^{\phantom{I}J},\Psi_J\right] \,.
      \end{split}
    \end{align}
    It is invariant under the SUSY transformations
    \begin{subequations}\label{eq:7dsusytrafos}
      \begin{align}
        \delta A_M&=\frac{\I}{2} \bar{\varepsilon}^I \Gamma_M \Psi_I\,,\\
        \delta B_i&=\frac{1}{2}\bar{\varepsilon}^I \left(\sigma_i\right)_I^{\phantom{I}J}\Psi_J\,,
        \\ 
        \delta\Psi_I &= -\frac{1}{4}F_{MN} \Gamma^{MN} \varepsilon_I +\frac{\I}{2} \Gamma^M D_M
        \left(B_i\sigma^i\right)_I^{\phantom{I}J} \varepsilon_J+ \frac{1}{4} \varepsilon^{ijk}
        \left[B_i,B_j\right] \left(\sigma_k\right)_I^{\phantom{I}J} \varepsilon_J\,.
      \end{align}
    \end{subequations}
    Here the transformation parameter is again a symplectic Majorana spinor $\varepsilon_I$.

    When checking the invariance of the Lagrangeans~(\ref{eq:L10}) and (\ref{eq:Lsym7d}), the only 
    pieces which do not cancel immediately are quartic expressions in the fermions, $\sim
    \Xi\Xi\,\Xi\epsilon$ and $\sim\Psi\Psi\,\Psi\epsilon$, respectively.  These can be seen to vanish 
    using Fierz transformations and the symmetry properties of (symplectic) Majorana spinor products.
    We collect our conventions regarding seven-dimensional spinors in Appendix~\ref{app:sevenspinors}.

    \subsection{Four-Dimensional Degrees of Freedom}
      We will now reformulate the Lagrangean in terms of four-dimensional degrees of freedom. Thus,
      the manifest Lorentz symmetry is broken to $SO(1,6)\to SO(1,3)\times SO(3)\cong SO(1,3)\times
      SU(2)$, while the $R$~symmetry is untouched. In the bosonic sector, the vector splits into a
      four-dimensional vector $A_\mu$ and a triplet of ``gauge scalars'' $A_i$. The $SU(2)_R$
      triplet $B_i$ just carries over\footnote{The $A_i$ and $B_i$ are triplets under different
        copies of $SU(2)$. To avoid excessive notation and for later convenience, we use the same
        indices.}. 

      The fermionic sector requires
      more work. From the $\Gamma$ matrices in Appendix~\ref{app:sevenspinors}, we see that the
      four-dimensional chirality matrix is
      \begin{align}
        \Gamma_{\!*}&=\I \Gamma^0\Gamma^1 \Gamma^2 \Gamma^3=\begin{pmatrix} -\mathbbm{1}
            &&&\\ 
            & \mathbbm{1} &&\\ && - \mathbbm{1} & \\ &&& \mathbbm{1}\end{pmatrix}\,.          
      \end{align}
      The charge conjugation matrix is
      \begin{align}
        B&=\Gamma^2\Gamma^5  =\begin{pmatrix} \text{\Large 0}&\begin{matrix} 0&
            -\epsilon_{\alpha\beta}\\ 
            -\epsilon^{\alphad\betad} & 0\end{matrix}\\
          \begin{matrix}  0& \epsilon_{\alpha\beta} \\ \epsilon^{\alphad\betad}
            &0\end{matrix} &\text{\Large 0}
        \end{pmatrix} \,.
      \end{align}
      Hence we can decompose the pair of symplectic Majorana gauginos in terms of four Weyl spinors
      as  follows:
      \begin{align}\label{eq:weyldecomposition}
        \Psi_1&=\begin{pmatrix}\lambda_{1\alpha}\\ \lambdab_2^\alphad \\ \lambda_{3\alpha}\\
          \lambdab_4^\alphad\end{pmatrix}\,, &
        \Psi_2&=\begin{pmatrix} -\lambda_{4\alpha}\\ -\lambdab_3^\alphad \\ \lambda_{2\alpha}\\
          \lambdab_1^\alphad\end{pmatrix}\,.
      \end{align}
      Similarly, the SUSY transformation parameter $\varepsilon_I$ can be decomposed into four Weyl
      spinors $\epsilon_1$ to $\epsilon_4$ and their conjugates. A direct computation shows that
      under the $R$~symmetry, the $\lambda_r$ arrange into doublets
      \begin{align}
        &\begin{pmatrix} \lambda_1\\\lambda_4\end{pmatrix}\,, & &\begin{pmatrix}
          \lambda_2\\\lambda_3\end{pmatrix}\,, 
      \end{align}
      while under the geometric $SU(2)$ we have doublets 
      \begin{align}
        &\begin{pmatrix} \lambda_1\\\lambda_3\end{pmatrix}\,, & &\begin{pmatrix}
          \lambda_2\\\lambda_4\end{pmatrix}\,.
      \end{align}
      Thus, the $\lambda_r$ transform as a
      $\left(\boldsymbol{2},\boldsymbol{2}\right)=\boldsymbol{4}$ under $SU(2)\times SU(2)\cong
      SO(4)$, while the scalar triplets $A_i\sim\left(\boldsymbol{3},\boldsymbol{1}\right)$ and
      $B_i\sim\left(\boldsymbol{1},\boldsymbol{3}\right)$ can be represented as (anti)-selfdual
      two-index tensors of $SO(4)$.

      In full dimensional reduction, the theory would obtain an $SU(4)$ $R$~symmetry as enhancement
      of the $SO(4)$. The spinors simply lift to a $\boldsymbol{4}$ of $SU(4)$ \cite{Sohnius:1980xd}. The
      scalars then become a $\boldsymbol{6}$, satisfying a reality condition (this is consistent
      because the $\boldsymbol{6}\sim\text{\tiny $\yng(1,1)$}$ is a self-conjugate representation,
      the condition being $\phi_{ij}=\frac{1}{2} \epsilon_{ijkl}\phib^{kl}$). This condition
      rules out a further enhancement of the $R$~symmetry to $U(4)$. Here, however, the scalar
      sector prohibits such an enhancement since the $A_i$ and $B_i$ are genuinely different, so
      indeed $SO(4)$ is the largest admissible $R$~symmetry group.

      \bigskip

      In Appendix~\ref{app:susytrafos} we have collected the supersymmetry transformations
      expressed explicitly in terms of four-dimensional quantities. For the superfield
      formulation, we have to single out one particular transformation parameter, which will break
      the manifest $R$~symmetry to $SO(3)$ which we identify with the diagonal $SU(2)$. In
      particular, the fermions decompose as 
      $\boldsymbol{4}\to\boldsymbol{1}\oplus\boldsymbol{3}$, while the scalars again form two
      triplets. The diagonal $SU(2)$ ensures gauge covariance, i.e.\ preservation of the two-triplet
      structure of the scalars without mixing the $A_i$ and $B_i$.

      From a practical point of view, we require the scalar components of the chiral multiplets to
      be of the form $\phi_i=A_i+ \I B_i$ and demand that $\delta\phi_i$ does not depend on
      $\bar{\epsilon}$.  This singles out the supersymmetry parameter choice
      $\epsilon_1=\epsilon_2=0$, $\epsilon_3=\epsilon_4\equiv \sqrt{2}\epsilon$.
      With this choice, the following fields transform as chiral multiplets:
      \begin{subequations}\label{eq:chiralcomp}
        \begin{align}
          \phi_1&=A_5+\I B_1 \,, & \psi_1&= \I \left(\lambda_1-\lambda_2\right)\,,\\
          \phi_2&=A_6+\I B_2 \,, & \psi_2&= -\left(\lambda_1+\lambda_2\right)\,,\\
          \phi_3&=A_7+\I B_3 \,, & \psi_3&= \I \left(\lambda_4-\lambda_3\right)\,.
        \end{align}
      \end{subequations}
      Explicitly, their transformation is
      \begin{align}
        \delta \phi_i&=\sqrt{2} \epsilon \psi_i\,, &\delta \psi_i&= -\sqrt{2}\I\left( \partial_\mu
          \phi_i -\partial_iA_\mu
        +\I\left[A_\mu,\phi_i\right]\right)
        \sigma^\mu \epsilonb -\sqrt{2} F_i \epsilon\,.
      \end{align}
      The extra terms in $\delta \psi_i$ make the right-hand side gauge covariant: The bracket is
      just $F_{\mu i} +\I D_\mu B_i$. The multiplets still are on-shell, i.e.\ the auxiliary fields
      are fixed to be
      \begin{align}\label{eq:auxCF}
        F_i&=-\frac{1}{2}\varepsilon_{ijk} \left(F_{jk} +2\I D_j B_k -\I
          \left[B_j,B_k\right]\right)\,. 
      \end{align}
      Here the expression in brackets is reminiscent of the field strength of the complex internal
      gauge field $\phi_i$.

      The remaining fields $A_\mu$ and $\chi=\lambda_3+\lambda_4$ form a vector multiplet, 
      \begin{align}\label{eq:vectorcomp}
        \delta A_\mu &= -\frac{\I}{\sqrt{2}} \left(\epsilon\sigma_\mu \chib -\chi\sigma_\mu
          \epsilonb\right) \,, & \delta \chi&= -\sqrt{2} F_{\mu\nu}\sigma^{\mu\nu} \epsilon
        +\sqrt{2}\I D \,\epsilon\,,
      \end{align}
      where the auxiliary field is $D=D_i B_i$. 
      
      Observe that the supersymmetry and gauge transformations mix:
      As usual for the vector multiplet, the supersymmetry transformations have to be accompanied
      by a gauge transformation which reestablishes Wess--Zumino gauge. This implies that they
      close only up to a gauge
      transformation,
      \begin{align}
        \left[\delta_\epsilon,\delta_\eta\right]A_\mu &= -2\I \left(\epsilon\sigma^\nu\etab
          -\eta\sigma^\nu\epsilon\right) \partial_\nu A_\mu -\delta_\text{gauge}\,.
      \end{align}
      Here $\delta_\text{gauge}$ is a transformation with the field-dependent parameter
      \begin{align}\label{eq:compensatinggaugetrafo}
        \I \left(\epsilon\sigma^\mu\etab -\eta\sigma^\mu\epsilon\right)A_\mu\,.
      \end{align}
      Now the same phenomenon appears for the chiral multiplets, which have a gauge condition
      imposed on them: The real part of the scalar component transforms inhomogeneously while the
      rest are tensors. This condition is violated by simple admixtures of $\phi_i\epsilon$ to
      $\psi_i$, and the violation needs to be compensated by a suitable gauge transformation which leads
      to~(\ref{eq:compensatinggaugetrafo}).

  \section{Superfield Lagrangean\label{sec:sf}}
    In this section we express the component theory of Section~\ref{sec:comp} in terms of
    $\mathcal{N}=1$ superfields. Our conventions regarding Weyl spinors and van der Waerden notation
    are given in Appendix~\ref{app:weylspinors}. We first specify the embedding of the fields into
    superfields. This will enable us to define a covariant extra-dimensional derivative, which in turn is crucial to
    formulate the Lagrangean.
    
    \subsection{Field Embedding}
      We embed the chiral multiplets~(\ref{eq:chiralcomp}) into chiral superfields $\Phi_i$, while
      the vector multiplet~(\ref{eq:vectorcomp}) forms a vector superfield in WZ gauge,
      \begin{align}
        V&= -\theta \sigma^\mu \thetab A_\mu +\frac{1}{\sqrt{2}}\theta^2\thetab\chib
        +\frac{1}{\sqrt{2}} \thetab^2 \theta\chi +\frac{1}{2} \theta^4 D\,,\\
        \Phi_i &= \phi_i +\sqrt{2}\I\theta\psi_i +\theta^2 F_i\,.
      \end{align}
      A gauge transformation now takes a complete chiral multiplet $\Lambda$ as parameter.
      (WZ gauge is preserved for $\Lambda$ having only a real scalar component.) The
      presence of the vector field components in $\phi_i$ implies that the superfields transform as 
      \begin{subequations}
        \begin{align}
          \Phi_i&\longrightarrow e^{-\I\Lambda} \left(\Phi_i-\I\partial_i\right) e^{\I\Lambda}\,,&
          e^{2V} &\longrightarrow e^{-\I\Lambdab} e^{2V} e^{\I\Lambda}\,.
        \end{align}
      \end{subequations}
      We define a supersymmetric covariant derivative in the extra dimensions, 
      \begin{align}
        \nabla_i&=\partial_i +\I \Phi_i\,,
      \end{align}
      which transforms as $\nabla_i\to e^{-\I\Lambda}\nabla_ie^{\I\Lambda}$. Here it is implied that $\Phi_i$ acts
      according to the representation of the field it is applied to.  In particular, we have
      \begin{align}\label{eq:nablaV}
        \nabla_i e^{2V}&= \partial_i e^{2V} +\I \Phib_i e^{2V} - \I e^{2V} \Phi_i
      \end{align}
      for the adjoint vector superfield.

    \subsection{Lagrangean}
      The Lagrangean contains three pieces: The usual gauge kinetic term, a kinetic term for the
      chiral superfields and a superpotential-like term. All three are by themselves invariant under
      $\mathcal{N}=1$ supersymmetry and gauge symmetry. Their relative coefficients are determined
      by higher-dimensional Lorentz invariance.
      
      \medskip
      
      There are two field-strength-like superfields that can be constructed from $V$ and $\Phi$. The
      first is the usual chiral field strength \mbox{$W_\alpha=-\frac{1}{4} \Db^2 e^{-2V}D_\alpha  e^{2
          V}$}, which gives rise to the action 
      \begin{align}
        \frac{1}{16}\int\D^2\theta\, W^\alpha W_\alpha +\text{H.c.}&=-\frac{1}{4}
        F_{\mu\nu}F^{\mu\nu} - \frac{\I}{2} \chi\sigma^\mu D_\mu \chib +\frac{1}{2}D^2\,.
      \end{align}

      The second piece contains extra-dimensional derivatives acting on $V$.
      Equation~(\ref{eq:nablaV}) implies that the simplest covariantly transforming object is
      \cite{Hebecker:2001ke} 
      \begin{align}
        Z_i&= e^{-2V} \nabla_i e^{2V}\,.
      \end{align}
      It is neither chiral nor real. Under gauge transformations, this field transforms as $Z_i\to
      e^{-\I\Lambda} Z_i e^{\I \Lambda}$. Note that this implies that the components of $Z_i$ are
      not tensors even under WZ~gauge preserving transformations (unless the gauge group is Abelian,
      in which case $Z_i$ is gauge invariant), for which the parameter is (in a
      non-chiral superfield representation)
      \begin{align}
        \Lambda &=\lambda +\I\theta \sigma^\mu\thetab \partial_\mu \lambda +\frac{1}{4}\theta^4
        \Box\lambda \,,
      \end{align}
      where $\lambda$ is real. Here the variation of $Z_i$ is
      \begin{align}
        \delta Z_i &= \I \left[Z_i,\Lambda\right]= \I \left[Z_i,\lambda\right] +\I\left[Z_i,\I\theta
          \sigma^\mu\thetab \partial_\mu \lambda +\frac{1}{4}\theta^4 \Box\lambda\right] \,.
      \end{align}
      The second term shows that the higher components ($\theta\thetab$ and higher) transform
      inhomogeneously. $Z_i$ is not Hermitean, but
      satisfies $Z_i^\dagger=e^{2V} Z_i e^{-2V}$, such that 
      the lowest-dimensional gauge-invariant term that can be formed, $\tr Z_i Z_i$ is
      Hermitean\footnote{Alternatively, one could define a real field $\widetilde{Z}_i=e^V Z_i
        e^{-V}$, for which $\tr \widetilde{Z}_i \widetilde{Z}_i=\tr Z_i Z_i$.}. 
      This is the second piece of the action,
      \begin{align}
        \begin{split}
          \frac{1}{4} \int \D^4\theta\, \tr Z_i Z_i&=-\frac{1}{2} F_{\mu i} F^{\mu i} -\frac{1}{2}
          D_\mu B_i D^\mu B_i - D D_i B_i +2 F_i \Fb_i\\
          &\quad-\frac{\I}{2} \psi_i\sigma^\mu D_\mu \psib_i-\frac{1}{2} \chi D_i
          \psi_i -\frac{1}{2}\chib D_i \psib_i\\
          &\quad +\frac{1}{2}\chi \left[B_i,\psi_i\right] -
          \frac{1}{2}\chib\left[B_i,\psib_i\right] \,. 
        \end{split}
      \end{align}

      \medskip
      
      The final piece contributes $F_{ij} F^{ij}$ and related terms via the auxiliary fields $F_i$.
      It is given as a Chern--Simons-like superpotential contribution,
      \begin{align}
        W&=\varepsilon_{ijk} \Phi_i\left(\partial_j \Phi_k
            +\frac{\I}{3}\left[\Phi_j,\Phi_k\right] \right)\,,
      \end{align}
      which gives rise to
      \begin{align}
        \begin{split}
          \frac{1}{4} \int \D^2\theta\, W
          +\text{H.c.}&=\frac{1}{4}
          \varepsilon_{ijk} F_i \left(F_{jk} + 2\I D_j B_k- \I \left[B_j,B_k\right]\right)\\
          &\quad +\frac{1}{4} \varepsilon_{ijk} \psi_i D_j \psi_k -\frac{1}{4} \varepsilon_{ijk}
          \psi_i \left[B_j,\psi_k\right]  +\text{H.c.}\\
        \end{split}
      \end{align}
      The left-hand side is not obviously gauge invariant. Rather, the superpotential transforms as
      \begin{align}\label{eq:deltaW}
        \delta W
        &=\frac{1}{3} \varepsilon_{ijk}
        \left(e^{-\I\Lambda}\partial_i e^{\I\Lambda}\right) \left(e^{-\I\Lambda}\partial_j
          e^{\I\Lambda}\right) \left(e^{-\I\Lambda}\partial_k e^{\I\Lambda}\right) \,.
      \end{align}
      However, we will now argue that the action is still gauge invariant: 
      First note that $\delta W$ depends only on the gauge transformation parameter
      $\Lambda$, but not on the fields $V$ or $\Phi_i$. Furthermore, $\delta W$ vanishes under the
      $\D^2\theta$ integral if $\Lambda$ contains only a scalar component, which includes WZ
      preserving gauge transformations, but also transformations which endow $V$ with a scalar
      ($\theta=\thetab=0$) component. 

      Second, we 
      recognise $\delta W$ as the ``winding number density''\cite{Jackiw:1976pf}, so the integral
      $\int \D^3 y\,\delta W=\cons \cdot n$ gives the winding number of the gauge
      transformation\footnote{This is the winding number around the internal space in the
        phenomenologically interesting case of three compact dimensions. For noncompact $y$
        directions, one has to assume suitable boundary conditions for the transformation as
        $\left|y\right|\to\infty$.}. 
      It is characterised by the third homotopy group, which is $\pi_3(G)=\mathbbm{Z}$ for all
      compact simple groups. 

      Third, 
      the integral $\int \D^3 y\,\delta 
      W=z+\theta \zeta+\theta^2 F_Z$ is a bona fide chiral superfield in four dimensions, since the $x$ and $\theta$
      dependence is untouched by the internal-space integral. Hence,
      under supersymmetry transformations its scalar component should transform as $\delta
      z=\epsilon\zeta$. However, this transformation together with the quantisation condition
      $z=\cons \cdot n$ implies that $\zeta=0$, and thus the transformation $\delta \zeta\sim F_Z
      \epsilon$ in turn requires $F_Z=0$. Hence $\int \D^2\theta\, \delta W=F_Z=0$, and the action is
      gauge invariant for any gauge transformation parameter $\Lambda$. 
      
      \medskip

      Altogether, we have the following Lagrangean:
      \begin{align}
        \begin{split}
          \mathscr{L}_\text{SF}&=\frac{1}{4} \int \D^4\theta\, \tr Z_i Z_i\\
          &\quad +
          \left[\frac{1}{16}\int\D^2\theta\, W^\alpha W_\alpha +\frac{1}{4} \int \D^2\theta \,
            \varepsilon_{ijk} \Phi_i\left(\partial_j \Phi_k +\frac{\I}{3}\left[\Phi_j,\Phi_k\right]
            \right) +\text{H.c.}\right] \\
          &=-\frac{1}{4}F_{\mu\nu}F^{\mu\nu} -\frac{1}{2} F_{\mu i} F^{\mu i}-\frac{1}{2} D_\mu B_i
          D^\mu B_i    +\frac{1}{2} D^2 -  D D_i B_i \\
          &\quad +\frac{1}{2} F_i \Fb_i +\left[\frac{1}{4} F_i \varepsilon_{ijk}\left(F_{jk} + 2\I
               D_j B_k- \I \left[B_j,B_k\right]\right) +\text{H.c.}\right]\\
          &\quad -\frac{\I}{2} \chi\sigma^\mu D_\mu\chib -\frac{\I}{2}\psi_i\sigma^\mu D_\mu \psib_i\\
          &\quad -\frac{1}{2}\left[\chi\left(D_i\psi_i-\left[B_i,\psi_i\right]\right)
            -\frac{1}{2} \varepsilon_{ijk} \psi_i\left(D_j
              \psi_k-\left[B_j,\psi_k\right]\right)+\text{H.c.} \right]\,. 
        \end{split}
      \end{align}
      Eliminating the auxiliary fields by their equations of motion, 
      \begin{subequations}\label{eq:auxSF}
        \begin{align}
          F_i&= -\frac{1}{2} \varepsilon_{ijk} \left(F_{jk} + 2\I D_j B_k- \I
            \left[B_j,B_k\right]\right) \,,\\
          D&= D_i B_i\,,
        \end{align}
      \end{subequations}
      we obtain the final expression
      \begin{align}
        \begin{split}
          \mathscr{L}_\text{SF}&= -\frac{1}{4}F_{MN} F^{MN} -\frac{1}{2} D_M B_iD^M B_i
          +\frac{1}{4}\left[B_i,B_j\right]  \left[B_i,B_j\right] \\
           &\quad -\frac{\I}{2} \chi\sigma^\mu D_\mu\chib -\frac{\I}{2}\psi_i\sigma^\mu D_\mu \psib_i\\
          &\quad -\frac{1}{2} \left[\chi\left(D_i\psi_i-\left[B_i,\psi_i\right]\right)
            -\frac{1}{2} \varepsilon_{ijk} \psi_i\left(D_j
              \psi_k-\left[B_j,\psi_k\right]\right)+\text{H.c.} \right]\,.
        \end{split}
      \end{align}
      This reproduces the original Lagrangean~(\ref{eq:Lsym7d}) when expressed in four-dimensional
      degrees of freedom. Note also that the auxiliary field expressions~(\ref{eq:auxSF})
      and~(\ref{eq:auxCF}) match. 

      All three pieces of the Lagrangean are by themselves 4D Lorentz invariant, $\mathcal{N}=1$
      supersymmetric and gauge invariant.
      Their relative coefficients are fixed by the requirement of seven-dimensional Lorentz
      symmetry, which, together with the manifest supersymmetry, enforces $\mathcal{N}=4$
      supersymmetry. Note that, in particular, the action does not contain explicit factors of $V$
      and consequently does not require a Wess--Zumino--Witten-like term
      \cite{Marcus:1983wb,ArkaniHamed:2001tb} to ensure gauge invariance under transformations not
      preserving  WZ gauge.

  \section{Conclusions and Outlook\label{sec:conclusions}}
    For SUSY model building it is rather convenient to have a simple superfield formulation. However, for
    higher-dimensional supersymmetric gauge theories, these formulations are often rather cumbersome, because of the 
    nontrivial interplay of supersymmetry and gauge symmetry. In this paper we have shown that the simple covariant
    formulation of \cite{Hebecker:2001ke} cannot be generalised to arbitrary dimensions, but only to the case of $D=7$.
    The origin of this fact is the combination of extra-dimensional Lorentz and $R$ symmetries, which requires an equal
    number of scalars and higher-dimensional vector components to form chiral multiplets.

    Furthermore, we have presented a gauge covariant superfield description of seven-dimensional super-Yang--Mills
    theory. The action contains the usual gauge kinetic term, a K\"ahler potential term and a superpotential. All three
    terms are by themselves $\mathcal{N}=1$ supersymmetric, 4D Lorentz invariant and invariant under superfield gauge
    transformations, including those that do not preserve WZ gauge. In particular, the vector superfield does not appear
    explicitly, and hence there is no need for a WZW-like term.  The $R$ symmetry argument alluded to above does
    additionally suggest that the six-dimensional $\mathcal{N}=4$ (maximal) super-Yang--Mills theory has a similar
    formulation, which can be obtained from the seven-dimensional case by reduction. 
    
    Seven-dimensional field theories can be studied in their own right. However, they also naturally appear in the
    context of type~IIA string theory with D6~branes compactified on a Calabi--Yau, or of M-theory on $G_2$
    manifolds with ADE singularities. The formalism presented here finds a natural application in
    this setup. Since $\mathcal{N}=1$ supersymmetry is manifest, and the coordinates are naturally split between the
    internal and Minkowski space, one can easily treat intersecting branes, which lead to a (possibly spontaneously
    broken) $\mathcal{N}=1$ supersymmetric theory in four dimensions.  The coupling of localised matter on the
    intersection to the gauge fields on the brane are directly apparent. Furthermore, the formalism allows for a
    systematic study of higher-dimensional operators in a supersymmetric fashion.

  \subsection*{Acknowledgements}
    We are grateful to Dmitry Belyaev, Wilfried Buchm\"uller,  Stefan Groot Nibbelink and Arthur Hebecker for helpful
    discussions and to Michael Blaszczyk and Fabian R\"uhle for comments on the manuscript.
    
    This work was partially supported by the SFB-Transregio TR33 "The Dark Universe" (Deutsche
    Forschungsgemeinschaft) and the European Union 7th network program "Unification in the LHC era"
    (PITN-GA-2009-237920). 

  \pagebreak
  \appendix 
  \section{Spinor Conventions\label{app:conventions}}
    \subsection{Generalities\label{app:gammas}}
      We use the metric $\eta=\diag\!\left(-,+,\dotsc,+\right)$. The $\Gamma$ matrices satisfy the
      algebra $\left\{\Gamma^M,\Gamma^N\right\}=2 \eta^{MN} \mathbbm{1}$. Hence, $\Gamma^0$ is
      anti-Hermitean, while the rest is Hermitean. 

      We denote seven-dimensional indices by $M,N,\dotsc=0,1,2,3,4,5,6$ and four-dimensional ones by
      $\mu,\nu,\dotsc=0,1,2,3$. Indices in the extra dimensions are denoted by $i,j,\dotsc=1,2,3$ and are raised
      and lowered with $\delta_{ij}$. 

    \subsection{Seven-Dimensional Spinors\label{app:sevenspinors}}
      For the reduction to four dimensions, we use the following explicit representation:
      \begin{align}\label{eq:gammas7}
        \Gamma^\mu&=\mathbbm{1}\otimes \gamma^\mu \,, & \Gamma^{3+i}&=\sigma^i\otimes \gamma^5\,,
      \end{align}
      where the $4\times 4$ matrices $\gamma^\mu$ are given by
      \begin{align}
        \gamma^0&=\begin{pmatrix}0&\mathbbm{1}\\-\mathbbm{1} &0\end{pmatrix}\,,
        &\gamma^i&=\begin{pmatrix}0&\sigma^i\\\sigma^i &0\end{pmatrix}\,,
      \end{align}
      and the $\sigma^i$ are the Pauli matrices. $\gamma^5$ is the product
      \begin{align}
        \gamma^5&=\I\gamma^0\gamma^1\gamma^2\gamma^3=\begin{pmatrix}-\mathbbm{1}&0\\0&\mathbbm{1} 
        \end{pmatrix} \,.
      \end{align}
      
      The gaugino satisfies a symplectic Majorana condition. It can be phrased in terms of the
      charge conjugation matrix $C$ which generates transpositions, or in terms of $B=-\Gamma^0 C$
      which generates complex conjugation,      
      \begin{align}
        C \Gamma_M C^{-1}&=-\Gamma_M^T\,, & B \Gamma_M B^{-1}&=\Gamma_M^*\,,
      \end{align}
      and reads in four equivalent formulations (in seven dimensions $BB^*=-\mathbbm{1}$)
      \begin{align}
        \Psi_I&=\varepsilon_{IJ} C\left(\Psib^T\right)^J\,, & \Psib^I&=-\varepsilon^{IJ}\Psi_J^T
        C\,,\\   
        \Psi_I&=\varepsilon_{IJ} B \left(\Psi^*\right)^J\,, &
        \left(\Psi^*\right)^J &=\varepsilon^{IJ}B^* \Psi_J\,. 
      \end{align}
      Explicitly, we have
      \begin{align}\label{eq:C7explicit}
        C&=\Gamma^0\Gamma^2\Gamma^5=\begin{pmatrix} \text{\Large 0}&\begin{matrix} \varepsilon& 0\\
              0& \varepsilon\end{matrix}\\
            \begin{matrix}  -\varepsilon &0 \\ 0& -\varepsilon\end{matrix} &\text{\Large 0}
          \end{pmatrix}\,, &B&=\Gamma^2\Gamma^5  =\begin{pmatrix} \text{\Large 0}&\begin{matrix} 0&
              -\varepsilon\\ 
              \varepsilon& 0\end{matrix}\\
            \begin{matrix}  0& \varepsilon \\ -\varepsilon &0\end{matrix} &\text{\Large 0}
          \end{pmatrix}\,.
      \end{align}

    \subsection{Weyl Spinors\label{app:weylspinors}}
      Our conventions regarding four-dimensional Weyl spinors are similar to~\cite{Sohnius:1985qm}, mainly
      differing by the different choice of metric. We denote
      left-handed (right-handed) Weyl spinors by undotted (dotted) indices. They are raised and
      lowered with the $\epsilon$ symbol in the following way: 
      \begin{align}
        \psi^\alpha&= \epsilon^{\alpha\beta}\psi_\beta\,, &\psi_\alpha&= \psi^\beta
        \epsilon_{\beta\alpha}\,,\\
        \chib^\alphad&= \chib_\betad \epsilon^{\betad \alphad}\,, &\chib_\alphad&=
        \epsilon_{\alphad\betad} \chib^\betad\,,
      \end{align}
      where $\epsilon$ is defined as
      \begin{align}
        \epsilon_{12}&=\epsilon^{12}=1\,, &\epsilon_{\dot1\dot2}&=\epsilon^{\dot1\dot2}=-1\,.
      \end{align}
      Note that this convention implies that
      $\epsilon^{\alpha\beta}\epsilon_{\beta\gamma}=-\delta^\alpha_\gamma$.  For spinor products,
      undotted indices are contracted top-down, dotted ones bottom up,
      \begin{align} 
        \psi \chi&=\psi^\alpha \chi_\alpha=\chi\psi\,, &\psib \chib&=\psib_\alphad
        \chib^\alphad=\chib\psib\,.
      \end{align} 
      Hermitean conjugation turns undotted into dotted indices and vice versa, 
      \begin{align}
        \left(\psi_\alpha\right)^\dagger&=\psib_\alphad\,, & 
        \left(\psi^\alpha\right)^\dagger&=\psib^\alphad\,, & \Rightarrow \quad 
        \left(\psi^\alpha\chi_\alpha\right)^\dagger&= \chib_\alphad\psib^\alphad= \chib\psib\,.
      \end{align}

      We define two sets of $\sigma$ matrices,
      $\sigma^\mu_{\alpha\alphad}=\left(\mathbbm{1},\sigma^i\right)$ and 
      $\left(\sigmat^\mu\right)^{\alphad\alpha}=\left(-\mathbbm{1},\sigma^i\right)$. They are
      related by raising indices, however, there is an extra minus sign,
      \begin{align}
        \epsilon^{\alpha\beta} \sigma^\mu_{\beta\betad}\epsilon^{\betad\alphad}&=-
        \left(\sigmat^\mu\right)^{\alphad \alpha}\,,& \epsilon_{\alphad\betad}
        \left(\sigmat^\mu\right)^{\betad\beta} \epsilon_{\beta\alpha}&=- \sigma^\mu_{\alpha
          \alphad}\,.
      \end{align}
      These conventions impose the following index structure on the $\Gamma$s and products
      thereof, repeating in each $4\times 4$ block: 
      \begin{align}
        \Gamma^{M\dotsm P}&=\begin{pmatrix}
          \begin{matrix}A_\alpha^{\phantom{\alpha}\beta} & B_{\alpha\betad} \\
            C^{\alphad\beta} & D^\alphad_{\phantom{\alphad}\betad}
          \end{matrix} & \text{\Large *}\\
          \text{\Large *} &\text{\Large *}
        \end{pmatrix}\,.
      \end{align}
      However, this does not apply to the charge conjugation matrix and the $\Gamma^0$ used in
      defining $\Psib=\Psi^\dagger \Gamma^0$, as these are intertwiners between different
      representations. In particular, we have
      \begin{align}
        B&=\Gamma^2\Gamma^5  =\begin{pmatrix} \text{\Large 0}&\begin{matrix} 0&
            -\epsilon_{\alpha\beta}\\ 
            -\epsilon^{\alphad\betad} & 0\end{matrix}\\
          \begin{matrix}  0& \epsilon_{\alpha\beta} \\ \epsilon^{\alphad\betad}
            &0\end{matrix} &\text{\Large 0} 
        \end{pmatrix}\,.
      \end{align}
      Furthermore, note that the $\epsilon$ symbols act in the wrong way (i.e.\ undotted bottom-up,
      dotted top-down), so in defining symplectic Majorana spinors there is another minus sign.
      Taking this together with the four-dimensional chirality matrix 
      \begin{align}
        \Gamma_{\!\!*}&=\I\Gamma^0\Gamma^1\Gamma^2\Gamma^3=\begin{pmatrix}-\mathbbm{1}
          &&&\\&\mathbbm{1}&& \\&& -\mathbbm{1}&\\&&&\mathbbm{1}\end{pmatrix}\,,
      \end{align}
      we see that a pair of seven-dimensional symplectic Majorana spinors decomposes into left- and
      right handed Weyl spinors $\lambda_a$ as 
      \begin{align}\label{eq:weyldecomposition2}
        \Psi_1&=\begin{pmatrix}\lambda_{1\alpha}\\ \lambdab_2^\alphad \\ \lambda_{3\alpha}\\
          \lambdab_4^\alphad\end{pmatrix}\,, &
        \Psi_2&=\begin{pmatrix} -\lambda_{4\alpha}\\ -\lambdab_3^\alphad \\ \lambda_{2\alpha}\\
          \lambdab_1^\alphad\end{pmatrix}\,.
      \end{align}
      For convenience, we also list the barred versions:
      \begin{align}
        \Psib_1&=\left(-\lambda_2^\alpha,\lambdab_{1\alphad},
          -\lambda_4^\alpha,\lambdab_{3\alphad}\right) \,,  &
        \Psib_2&=\left(\lambda_3^\alpha,-\lambdab_{4\alphad},
          -\lambda_1^\alpha,\lambdab_{2\alphad}\right)\,.
      \end{align}
      
  \section{Reduction from Ten to Seven Dimensions\label{sec:10to7}}
    In this section we detail the reduction of the ten-dimensional Majorana--Weyl spinor $\Xi$ to
    a seven-dimensional symplectic Majorana spinor. For the purpose of this Appendix, we denote
    ten-dimensional indices by $\Mh,\Nh=0,\dotsc,9$, and ten-dimensional quantities by hats.

    The appearance of the symplectic reality condition can be understood as follows: As a Weyl
    spinor, $\Xi$ transforms in the 
    $\boldsymbol{16}$  of $SO(1,9)$. (Recall that $SO(1,9)$ has two spinor representations,
    $\boldsymbol{16}$ and $\boldsymbol{16}'$, which are self-conjugate under charge conjugation.) Hence one can impose
    an additional Majorana condition, $\Xi^c=\Xi$, to reduce the number of real degrees of freedom to
    16. In the reduction to seven dimensions, i.e.\ in $SO(1,9)\to SO(1,6)\times SO(3)\cong
    SO(1,6)\times SU(2)$, the spinor decomposes as
    $\boldsymbol{16}\to\left(\boldsymbol{8},\boldsymbol{2}\right)$. Here the $\boldsymbol{8}$ is the
    spinor representation of $SO(1,6)$. The Majorana condition in ten dimensions translates into a
    symplectic condition acting on the $SU(2)$ doublet index. 

    To make this explicit, assume a set $\Gamma^M$, $M=0,\dotsc,6$, of seven-dimensional (but
    $8\times 8$)  $\Gamma$ matrices. Then a convenient set of $\Gamma$ matrices in ten dimensions is
    given by 
    \begin{align}
      \begin{aligned}
        \Gammah^M&= \sigma_3\otimes\sigma_3\otimes \Gamma^M\,,  \mspace{120mu}   
        &\Gammah^7&=\sigma_2\otimes\mathbbm{1}\otimes\mathbbm{1}_8\,,\\        
        \Gammah^8&=-\sigma_1\otimes\mathbbm{1}\otimes\mathbbm{1}_8\,,
        &\Gammah^9&=\sigma_3\otimes\sigma_1\otimes\mathbbm{1}_8\,.
      \end{aligned}
    \end{align}
    Here $\mathbbm{1}$ without subscript denotes the $2\times2$ unit matrix. This set is convenient
    because the decomposition of the 32-component ten-dimensional spinor into eight-component
    seven-dimensional spinors is directly apparent. Furthermore, the
    relation~(\ref{eq:gamasigme10-7}) allows for easy identification of the ten-dimensional fields
    with the seven-dimensional ones. 

    Let us now impose the Majorana--Weyl nature of the gaugino. The chirality matrix is
    \begin{align}
      \Gammah_{\!\!*}&=\Gammah^0 \dotsm\Gammah^9 =\sigma_3\otimes\sigma_2\otimes \mathbbm{1}_8 \,.
    \end{align}
    Here we assumed that $\Gamma^0\dotsc\Gamma^6=-\mathbbm{1}_8$, as happens for our
    choice~(\ref{eq:gammas7}). The only other possibility is
    $\Gamma^0\dotsc\Gamma^6=+\mathbbm{1}_8$, in which case we exchange what we call left- and
    right-handed spinors. The chirality condition $\Gammah_{\!\!*}\Xi=-\Xi$ then implies that $\Xi$
    is of the form
    \begin{align}
      \Xi&=\begin{pmatrix}\I \xi_1\\ \xi_1\\ \I \xi_2 \\ -\xi_2\end{pmatrix}\,,
    \end{align}
    with -- so far -- unconstrained eight-component spinors $\xi_{1,2}$.  

    The Majorana constraint involves the intertwiner with the transposed
    representation. Let $C$ be this matrix in seven dimensions (Eq.~(\ref{eq:C7explicit}) for our
    explicit case), such that $C\Gamma_MC^{-1}=-\Gamma_M^T$. In seven dimensions, this $C$ is
    symmetric and real\footnote{The reality and normalisation involve a choice of prefactor, since for
      any nonzero complex $\alpha$, $\alpha C$ intertwines just as well as $C$. However, apart from
      this rescaling, $C$ is unique.},
    i.e.\ $C=C^T=C^*$, such that $C=C^{-1}$. Then in ten dimensions, this intertwiner is
    \begin{align}
      \Ch&= \sigma_1\otimes\sigma_3\otimes C\,,
    \end{align}
    which is again real and symmetric and satisfies
    \begin{align}
      \Ch \Gammah_{\!\!\Mh}\Ch^{-1}&=\Gammah_{\!\!\Mh}^T\,.
    \end{align}
    The Majorana condition is thus 
    \begin{align}
      \Xi=\begin{pmatrix}\I \xi_1\\ \xi_1\\ \I \xi_2 \\ -\xi_2\end{pmatrix}&=\Xi^c=\Ch
      \overline{\Xi}^T=\begin{pmatrix}\I C\xib_2^T\\ C\xib_2^T\\ -\I C\xib_1^T \\
        C\xib_1^T\end{pmatrix} \,, 
    \end{align}
    where we can directly identify the symplectic Majorana condition on the eight-component spinors,
    \begin{align}
      \xi_1&=C\xib_2^T\,, & \xi_2&=-C\xib_1^T\,.
    \end{align}
    Note that $\Xi^c$ is still left-handed. One can check that in ten dimensions,
    $\left(\Xi^c\right)^c=\Xi$. Furthermore, note that indeed the Lorentz
    generators $\Sigma^{78}$, $\Sigma^{89}$ and $\Sigma^{97}$ generate an $SU(2)$ rotating $\xi_1$
    and $\xi_2$ into each other  (since $\Gammah^{7,8,9}$ act as the unit matrix on the
    eight-component spinors):
    \begin{align}
      \begin{aligned}
        \Sigma^{89}&:\begin{pmatrix}\xi_1\\\xi_2\end{pmatrix}\longrightarrow
        -\frac{1}{2}\begin{pmatrix}0& 1\\1&0        
        \end{pmatrix}\begin{pmatrix}\xi_1\\\xi_2\end{pmatrix}\,, &
        \Sigma^{97}&:\begin{pmatrix}\xi_1\\\xi_2\end{pmatrix}\longrightarrow
        -\frac{1}{2}\begin{pmatrix}0& -\I\\\I&0        
        \end{pmatrix}\begin{pmatrix}\xi_1\\\xi_2\end{pmatrix}\,, \\[2mm]
        \Sigma^{78}&:\begin{pmatrix}\xi_1\\\xi_2\end{pmatrix}\longrightarrow
        -\frac{1}{2}\begin{pmatrix}1& 0\\0&-1        
        \end{pmatrix}\begin{pmatrix}\xi_1\\\xi_2\end{pmatrix}\,.
      \end{aligned}
    \end{align}
   
    \medskip
    
    To identify the seven-dimensional fields, observe that the $\Gammah$s have been chosen such that
    the seven-dimensional $\Gamma$ matrices are already there, so the kinetic term directly
    descends. Furthermore, we have
    \begin{align}\label{eq:gamasigme10-7}
      \overline{\Xi}\,\Gammah^{7,8,9}\Xi &=-2\I \xib_I \sigma^{1,2,3}_{IJ} \xi_J\,.
    \end{align}
    We thus recover the Lagrangean~(\ref{eq:Lsym7d}) with  the identifications
    \begin{align}
      A_M&=\widehat{A}_M\,, & B_i&=\widehat{A}_{6+i}\,, & \Psi_I&= \sqrt{2}\,\xi_I\,.
    \end{align}

  \section{SUSY Transformations\label{app:susytrafos}}
      Here we list the SUSY variations, Eqns.~(\ref{eq:7dsusytrafos}), expressed in terms of
      four-dimensional degrees of freedom. In particular, we decompose the gaugino $\Psi_I$ and the
      transformation parameter $\varepsilon_I$ into Weyl spinors $\lambda_a$, $\epsilon_a$,
      $a=1,\dotsc,4$, according to Eq.~(\ref{eq:weyldecomposition}). Then we  get for the bosons: 
      \begin{subequations}
        \begin{align}
          \begin{split}
            \delta A_\mu &=\frac{\I}{2} \left[ \epsilonb_1 \sigmat_\mu \lambda_1 -
              \epsilon_1 \sigma_\mu\lambdab_1 +\epsilonb_2 \sigmat_\mu \lambda_2 -
              \epsilon_2 \sigma_\mu\lambdab_2\right.\\
              &\quad \left. \mspace{40mu}+\epsilonb_3 \sigmat_\mu \lambda_3 -
              \epsilon_3 \sigma_\mu\lambdab_3 +\epsilonb_4 \sigmat_\mu \lambda_4 -
              \epsilon_4 \sigma_\mu\lambdab_4 \right]\,,
          \end{split}\\[4mm]
          \delta A_4&= \frac{\I}{2} \left[\epsilonb_1\lambdab_4
            -\epsilon_1\lambda_4 -\epsilonb_2\lambdab_3+\epsilon_2\lambda_3
            +\epsilonb_3\lambdab_2 - \epsilon_3\lambda_2 -\epsilonb_4
            \lambdab_1 +\epsilon_4\lambda_1\right]\,,\\
          \delta A_5&= \frac{1}{2} \left[\epsilonb_1\lambdab_4
            +\epsilon_1\lambda_4 +\epsilonb_2\lambdab_3+\epsilon_2\lambda_3
            -\epsilonb_3\lambdab_2 - \epsilon_3\lambda_2 -\epsilonb_4
            \lambdab_1 -\epsilon_4\lambda_1\right]\,,\\
          \delta A_6 &= \frac{\I}{2} \left[\epsilonb_1\lambdab_2
            -\epsilon_1\lambda_2 -\epsilonb_2 \lambdab_1 + \epsilon_2 \lambda_1
            -\epsilonb_3\lambdab_4 +\epsilon_3\lambda_4 +\epsilonb_4\lambdab_3
            -\epsilon_4\lambda_3\right]\,,\\[4mm]
          \delta B_1 &= \frac{1}{2} \left[-\epsilonb_1\lambdab_3 -
            \epsilon_1\lambda_3 +\epsilonb_2\lambdab_4 +\epsilon_2 \lambda_4
            +\epsilonb_3\lambdab_1 + \epsilon_3\lambda_1 -\epsilonb_4\lambdab_2
            -\epsilon_4\lambda_2\right]\,,\\
          \delta B_2 &= \frac{\I}{2} \left[\epsilonb_1\lambdab_3 -
            \epsilon_1\lambda_3 +\epsilonb_2\lambdab_4 -\epsilon_2 \lambda_4
            -\epsilonb_3\lambdab_1 + \epsilon_3\lambda_1 -\epsilonb_4\lambdab_2
            +\epsilon_4\lambda_2\right]\,,\\
          \delta B_3&=\frac{1}{2} \left[\epsilonb_1\lambdab_2
            +\epsilon_1\lambda_2 -\epsilonb_2\lambdab_1 -\epsilon_2\lambda_1
            +\epsilonb_3\lambdab_4 +\epsilon_3\lambda_4 -\epsilonb_4\lambdab_3
            -\epsilon_4\lambda_3\right]\,.
        \end{align}
      \end{subequations}
      
      For the fermions we obtain{\small
      \begin{subequations}
        \begin{align}
          \begin{split}
            \delta \lambda_1&=-\frac{1}{2} \left(F_{\mu\nu}\sigma^{\mu\nu} +\I F_{45}
            \right)\epsilon_1 -\frac{1}{2} F_{\mu6}\sigma^\mu \epsilonb_2 -\frac{1}{2} 
            \left(F_{64}- \I F_{65}\right) \epsilon_3 -\frac{1}{2} \left(F_{\mu4}- \I F_{\mu5}\right)
            \sigma^\mu \epsilonb_4\\ 
            &\quad -\frac{\I}{2} D_\mu \left(B_1-\I
              B_2\right) \sigma^\mu \epsilonb_3 +\frac{\I}{2} D_\mu B_3
            \sigma^\mu \epsilonb_2 -\frac{\I}{2}\left(D_4-\I D_5\right)
            \left(B_1-\I B_2\right)\epsilon_2\\
            &\quad -\frac{\I}{2} \left(D_4-\I D_5\right) B_3
            \epsilon_3 +\frac{\I}{2} D_6\left(B_1-\I B_2\right) \epsilon_4
            -\frac{\I}{2} D_6 B_3 \epsilon_1\\
            &\quad +\frac{1}{2}\left[B_1,B_2\right] \epsilon_1 -\frac{\I}{2} \left[B_1-\I
              B_2, B_3\right] \epsilon_4\,,
          \end{split}\\*
          \begin{split}
            \delta \lambda_2&=-\frac{1}{2}  \left(F_{\mu\nu}\sigma^{\mu\nu} 
              -\I F_{45}\right) \epsilon_2 
            +\frac{1}{2} F_{\mu6}\sigma^\mu \epsilonb_1 
            -\frac{1}{2}\left(F_{64}+ \I F_{65}\right) \epsilon_4
            +\frac{1}{2} \left(F_{\mu4}+ \I F_{\mu5}\right) \sigma^\mu \epsilonb_3\\
            &\quad +\frac{\I}{2} D_\mu\left(B_1+\I B_2\right) \sigma^\mu \epsilonb_4
            -\frac{\I}{2} D_\mu B_3 \sigma^\mu \epsilonb_1 -\frac{\I}{2} \left(D_4+\I
              D_5\right) \left(B_1+\I B_2\right) \epsilon_1 \\
            &\quad -\frac{\I}{2} \left(D_4+\I
              D_5\right) B_3\epsilon_4 +\frac{\I}{2} D_6\left( B_1+\I B_2\right)\epsilon_3
            -\frac{\I}{2} D_6 B_3 \epsilon_2\\
            &\quad -\frac{1}{2}\left[B_1,B_2\right] \epsilon_2 -\frac{\I}{2}
            \left[B_1+\I B_2,B_3\right] \epsilon_3\,,
          \end{split}\\*
          \begin{split}
            \delta \lambda_3&=-\frac{1}{2}  \left(F_{\mu\nu}\sigma^{\mu\nu} 
              -\I F_{45} \right)\epsilon_3 
            +\frac{1}{2} F_{\mu6}\sigma^\mu \epsilonb_4 
            +\frac{1}{2}\left(F_{64}+ \I F_{65}\right) \epsilon_1
            -\frac{1}{2} \left(F_{\mu4}+ \I F_{\mu5}\right) \sigma^\mu \epsilonb_2\\
            &\quad +\frac{\I}{2} D_\mu \left(B_1-\I
              B_2\right) \sigma^\mu \epsilonb_1 +\frac{\I}{2} D_\mu B_3
            \sigma^\mu \epsilonb_4 +\frac{\I}{2}\left(D_4+\I D_5\right)
            \left(B_1-\I B_2\right)\epsilon_4\\
            &\quad -\frac{\I}{2} \left(D_4+\I D_5\right) B_3 \epsilon_1 +\frac{\I}{2}
            D_6\left(B_1-\I B_2\right) \epsilon_2 +\frac{\I}{2} D_6 B_3 \epsilon_3\\
            &\quad +\frac{1}{2}\left[B_1,B_2\right] \epsilon_3 +\frac{\I}{2} \left[B_1-\I
              B_2, B_3\right] \epsilon_2\,,
          \end{split}\\*
          \begin{split}
            \delta \lambda_4&=-\frac{1}{2}  \left(F_{\mu\nu}\sigma^{\mu\nu} +
              \I F_{45} \right)\epsilon_4 -\frac{1}{2} F_{\mu6}\sigma^\mu \epsilonb_3
            +\frac{1}{2}   \left(F_{64}- \I F_{65}\right) \epsilon_2 +\frac{1}{2} \left(F_{\mu4}- \I
              F_{\mu5}\right)  \sigma^\mu \epsilonb_1\\
            &\quad -\frac{\I}{2} D_\mu\left(B_1+\I  B_2\right) \sigma^\mu \epsilonb_2
            -\frac{\I}{2} D_\mu B_3 \sigma^\mu \epsilonb_3 +\frac{\I}{2} \left(D_4-\I
              D_5\right) \left(B_1+\I B_2\right) \epsilon_3 \\
            &\quad -\frac{\I}{2} \left(D_4-\I D_5\right) B_3\epsilon_2 +\frac{\I}{2} D_6\left(
              B_1+\I B_2\right)\epsilon_1  +\frac{\I}{2} D_6 B_3 \epsilon_4\\
            &\quad -\frac{1}{2}\left[B_1,B_2\right] \epsilon_4 +\frac{\I}{2}
            \left[B_1+\I B_2,B_3\right] \epsilon_1\,.
          \end{split}
        \end{align}
      \end{subequations}}

\end{document}